\documentclass[12pt]{iopart}
\expandafter\let\csname equation*\endcsname\relax
\expandafter\let\csname endequation*\endcsname\relax
\usepackage{amsmath}
\usepackage{bm}
\usepackage{graphicx}
\usepackage{graphics}
\usepackage{amsfonts}
\usepackage{amssymb}
\usepackage{makeidx}
\usepackage[colorlinks=true,citecolor=blue,linkcolor=blue,linktocpage=true,pagebackref=false]{hyperref}
\hypersetup{colorlinks=true,citecolor=blue,linkcolor=blue,filecolor=blue,urlcolor=blue}
\usepackage{color,soul} 
\usepackage{float}
\usepackage{subfigure}
\usepackage{ifthen}
\usepackage[normalem]{ulem}

\begin{document}

\title{Many-body density of states of bosonic and fermionic gases: a combinatorial approach}

\author{Carolyn Echter$^{1}$, Georg Maier$^1$, Juan-Diego Urbina$^1$, \\ 
Caio Lewenkopf$\,^2$ and Klaus Richter$^1$}
\address{
$^1$ Institut f\"ur Theoretische Physik, Universit\"at Regensburg, D-93040 Regensburg, Germany \\
$^2$ Instituto de F\'{\i}sica, Universidade Federal Fluminense, 24210-346 Niter\'oi, RJ, Brazil}
\ead{carolyn.echter@ur.de}

\begin{indented}
\item 
\item \today
\end{indented}

\begin{abstract}
We use a combinatorial approach to obtain exact expressions for the many-body density of states of fermionic and bosonic gases with equally spaced single-particle spectra. We identify a mapping that reveals a remarkable property, namely, fermionic and bosonic gases have the same many-body density of states, up to a shift corresponding to ground state energy. Additionally, we show that there is a regime, comprising the validity range of the Bethe approximation, where the many-body density of states becomes independent of the number of particles.
\end{abstract}

\section{Introduction}
\label{sec:intro}

Recent experiments with cold atoms \cite{Gross2017, Schafer2020} and the interest in many-body (MB) systems that involve high excitations, such as MB thermalization \cite{Srednicki1994, Deutsch1991, Rigol2008} and MB localization \cite{Oganesyan2007, Huse2015}, have rekindled the need for analytical estimates of the many-body density of states (MBDOS) and its fluctuations \cite{Comtet2007,Leboeuf2006, Lefevre2023}.
In particular, the celebrated Bethe estimate for the mean level density of fermionic systems in the mean-field regime has drawn much recent attention due to its ubiquitous presence  in systems with a holographic gravitational dual like SYK \cite{Garcia-Garcia2017} and also in low-dimensional dilaton gravity \cite{MISC:4}. 

For non-interacting quantum systems, determining the MBDOS translates into the combinatorial problem of counting the number of ways the single-particle (SP) spectral energies can add up to a given MB energy. 
While this problem is readily formulated, in general no exact analytical solution exists \cite{Leboeuf2005b}. Approximate results can be obtained using a partition function approach. 
Those include the mentioned Bethe formula for the level density of non-interacting fermionic MB systems \cite{Bethe1936,Bloch1954,BohrMottelson1998,QuirinJPA,Alhassid2021} and, more recently, its extensions to bosonic systems \cite{Comtet2007, Ribeiro2015}. 

For the particular case of MB systems with equally spaced SP energy levels, which is relevant as the statistical description of an ensemble of harmonic oscillators in one spatial dimension, an exact solution to the counting problem is known \cite{Auluck1946}. 
This result is stated and expanded on in Sec.~\ref{sec:combinatorial}. 
Remarkably, we find that in this model, there is a one-to-one correspondence between the bosonic and fermionic MBDOS. 
Consistently, it turns out that the combinatorial result reproduces the Bethe approximation for the MB harmonic oscillator within its range of validity and that this approximation is also valid for bosons, whereas the general Bethe formula does not immediately translate to the bosonic case. 
For systems with constant mean SP energy spacing, the Bethe formula predicts a particle number independent MBDOS. 
This coincides with our findings in the range of validity of the Bethe approximation.

The paper is organized as follows.
Sec.~\ref{sec:background} outlines the derivation and the range of validity of the Bethe approximation and of its bosonic analogue. 
Sec.~\ref{sec:combinatorial} is dedicated to discussing the combinatorial solution to the MBDOS in the case of equally spaced SP spectra. Eq.~\eqref{eq:7} summarizes the main result, and its derivation through Eq.~\eqref{eq:106} closes a gap in the existing literature. 
We show that the Bethe formula and its bosonic counterpart are recovered from the combinatorial results in a limit consistent with their validity range.
Further limits as well as prospects of generalization are discussed.
The conclusions are presented in Sec.~\ref{sec:conclusion}.

\section{Background}
\label{sec:background}

Let us consider a MB system of $N$ non-interacting particles. 
Let $\nu$ label its SP levels with energies $\epsilon_\nu$ and occupation numbers $n_\nu$, so that a MB configuration is characterized by the tuple $\left(n_\nu\right)$. 
Write
\begin{equation}
N_{\left(n_\nu\right)} = \sum_\nu n_\nu 
\qquad \mbox{and} \qquad
E_{\left(n_\nu\right)} = \sum_\nu n_\nu\epsilon_\nu
\end{equation}
for the number of particles and the MB energy corresponding to the configuration $\left(n_\nu\right)$ respectively. 
The MBDOS as a function of particle number $N$ and energy $E$ is given by
\begin{equation}
\rho(N,E) = \sum_{\left(n_\nu\right)}\delta\!\left(N-N_{\left(n_\nu\right)}\right)\delta\!\left(E-E_{\left(n_\nu\right)}\right).
\end{equation}

\subsection{Standard derivation of the MBDOS of a fermionic gas}
\label{sec:Bethe}

The Bethe formula can be derived, following Ref.~\cite{BohrMottelson1998}, by recognizing that the grand-canonical partition function is the Laplace transform of the MBDOS, namely,
\begin{equation}
Z(\alpha,\beta) = \sum_{\left(n_\nu\right)} \text{exp}\!\left(\alpha N_{\left(n_\nu\right)}-\beta E_{\left(n_\nu\right)}\right) 
= \int_0 ^\infty\!\!\text{d}N\int_0^\infty\!\!\text{d}{E}~\rho(N,E)\text{exp}(\alpha N - \beta E),
\end{equation}
where $\beta$ is inverse temperature and the dimensionless variable $\alpha = \beta\mu$ measures the chemical potential $\mu$ in units of $\beta^{-1}$. 
Consequently, the MBDOS can be obtained from the inverse Laplace transformation
\begin{align}
\label{eq:26}
\rho(N,E) & = 
 \left(\frac{1}{2\pi\mathrm{i}}\right)^2\int_{-\mathrm{i}\infty+\gamma}^{\mathrm{i}\infty+\gamma}\text{d}\alpha\int_{-\mathrm{i}\infty+\delta}^{\mathrm{i}\infty+\delta}\text{d}\beta~\text{exp}\!\left[\Phi(\alpha,\beta)\right],
\end{align}
where
\begin{equation}
\label{eq:entropy}
\Phi(\alpha,\beta) = -\alpha N + \beta E + {\rm log}~ Z(\alpha,\beta)
\end{equation}
is the entropy in units of the Boltzmann constant, 
$-\beta^{-1}\text{log}Z(\alpha,\beta)$ identifies with the grand canonical potential and $\gamma, \delta > 0$ are chosen in line with the inversion formula for the Laplace transform. 

For non-interacting fermions, $\text{log}~Z(\alpha,\beta)$ is readily written as \cite{Pathria2016book}
\begin{equation}
\label{eq:28}
\text{log}~Z(\alpha,\beta) = \int_0^\infty\!\text{d}\epsilon~g(\epsilon)\text{log}\!\left[1+\text{exp}(\alpha-\beta\epsilon)\right],
\end{equation}
where $g(\epsilon) = \sum_\nu\delta\!\left(\epsilon-\epsilon_\nu\right)$ is the system SP level density.
Assuming that the interval in which the integrand of Eq.~\eqref{eq:28} is notably different from zero is wide compared to the spacing of the SP levels, one may replace the SP level density by a smooth function of the energy, here also called $g$. 
If at the same time, this window is narrow compared to the scale over which $g$ varies, the problem satisfies the conditions of the Sommerfeld integral \cite{Pathria2016book}.
Hence, $\text{log}~Z(\alpha,\beta)$ can be written as 
\begin{equation}
\label{eq:Sommerfeld-expansion}
\text{log}~Z(\alpha,\beta) = \int_0^{\frac{\alpha}{\beta}}\!\text{d}\epsilon~g(\epsilon)(\alpha-\beta\epsilon) + \frac{\pi^2}{6\beta}g\!\left(\frac{\alpha}{\beta}\right) + \frac{7\pi^4}{360\beta^3}g''\!\left(\frac{\alpha}{\beta}\right) + \dots .
\end{equation}
Inserting Eq.~\eqref{eq:Sommerfeld-expansion} into Eq.~\eqref{eq:entropy} gives an approximate expression for the exponent in the inverse Laplace transform \eqref{eq:26}.
In a further approximation, the integral is evaluated using the saddle-point method to yield
\begin{equation}
\label{eq:33}
\rho(N,E) \approx \frac{\text{exp}\!\left(\Phi\!\left(\alpha_0,\beta_0\right)\right)}{2\pi\sqrt{\left|\text{det}\!\left(\Phi''\!\left(\alpha_0,\beta_0\right)\right)\right|}},
\end{equation}
where the (single) stationary point $\left(\alpha_0,\beta_0\right)$ is determined by
$\Phi'\!\left(\alpha_0,\beta_0\right) = 0$.

At the stationary-point, the leading order of Eq.~\eqref{eq:Sommerfeld-expansion} gives
\begin{equation}
\label{eq:35}
N = \int_0^{\frac{\alpha_0}{\beta_0}}\!\text{d}\epsilon~g(\epsilon) 
\qquad \mbox{and} \qquad
E = \int_0^{\frac{\alpha_0}{\beta_0}}\!\text{d}\epsilon~g(\epsilon)\epsilon + \frac{\pi^2}{6\beta_0^2}g\!\left(\frac{\alpha_0}{\beta_0}\right)\;.
\end{equation}
A comparison with the standard relations
\begin{equation}
\label{eq:36}
\int_0^{\epsilon_F}\!\text{d}\epsilon~g(\epsilon) = N 
\qquad \mbox{and} \qquad
\int_0^{\epsilon_F}\!\text{d}\epsilon~g(\epsilon)\epsilon = E-Q = E_{\rm GS}\,,
\end{equation}
where $\epsilon_F$ is the Fermi energy, $Q$ the excitation energy above the MB ground state and $E_{\rm GS}$ its ground states energy, allows one to identify (under the assumptions on $g$ which we make within the approximation, see the discussion of the range of validity below)
\begin{equation}
\label{eq:37}
\frac{\alpha_0}{\beta_0} = \epsilon_F 
\qquad \mbox{and} \qquad
\frac{\pi^2}{6\beta_0^2}g\!\left(\epsilon_F\right) = Q .
\end{equation}

By neglecting the derivatives of $g$, one can use the above relations to express
$\Phi\!\left(\alpha_0,\beta_0\right)$ and $\left|\text{det}\!\left(\Phi''\!\left(\alpha_0,\beta_0\right)\right)\right|$ in terms of  $\epsilon_F(N)$ and $Q(N,E)$ to obtain
\begin{equation}
\label{eq:38}
\rho(N,E) \approx \frac{1}{4\sqrt{3}Q}\text{exp}\!\left(\pi\sqrt{\frac{2}{3}g\!\left(\epsilon_F\right)Q}\right),
\end{equation}
which is the Bethe formula \cite{Bethe1936} for the MBDOS of fermionic gases.

Let us now discuss the range of validity of the Bethe formula, Eq.~\eqref{eq:38}.
The approximations involved include the replacement of the SP level density by a smooth function. 
At the stationary point, this approximation on the one hand necessitates that $\beta_0^{-1}g\!\left(\epsilon_F\right) \gg 1$
or, using Eq.~\eqref{eq:37}, equivalently
\begin{equation}
\label{eq:39}
g\!\left(\epsilon_F\right)Q \gg 1.
\end{equation}
Physically, this means that the excitation energy is required to be much larger than the mean level spacing at the Fermi energy, so that the system is not probed in its ground state. 
On the other hand, the condition $\beta^{-1}g'(\alpha/\beta) \ll g(\alpha/\beta)$ for the Sommerfeld expansion translates as
\begin{equation}
\label{eq:Bethe-req-2}
\frac{\left(g'\!\left(\epsilon_F\right)\right)^2Q}{\left(g\!\left(\epsilon_F\right)\right)^3} \ll 1.
\end{equation}
Furthermore, derivatives of $g$ were neglected, which by Eq.~\eqref{eq:Sommerfeld-expansion} requires
\begin{equation}
\label{eq:Bethe-req-3}
\frac{\left(g^{(2n)}\!\left(\epsilon_F\right)\right)^2Q^{2n+1}}{\left(g\!\left(\epsilon_F\right)\right)^{2n+1}} \ll 1
\end{equation}
for $n > 0$. 
Note that the range of validity of the saddle-point approximation is covered by Eq.~\eqref{eq:39} \cite{BohrMottelson1998}. 
Equations \eqref{eq:39} through \eqref{eq:Bethe-req-2} show that the different approximations compete in terms of the values of $Q$ for which they are valid. 
Importantly, this analysis does not capture the fact that the approach breaks down for very large values of $Q$, relative to the number of particles, corresponding to excitations of a large fraction of the Fermi sea.
For instance, for the case of constant SP level density, Eqs.~\eqref{eq:Bethe-req-2} and \eqref{eq:Bethe-req-3} would indicate that the approximations are always very accurate, but exact results obtained from the combinatorial analysis presented in Sec.~\ref{sec:combinatorial} show otherwise.

\subsection{Towards a bosonic analogue}
\label{sec:Bethe-bosonic}

The formulation of a bosonic analogue of the Bethe approximation poses an immediate problem. 
Namely, the derivation of Eq.~\eqref{eq:38} relies on the existence of a characteristic energy scale set by the Fermi energy, around which the SP level density is expanded. 
There is no evident analogue of such an energy scale in the bosonic case. 
Mathematically, this problem is reflected by the infrared divergence of the bosonic grand-canonical partition function
\begin{equation}
\label{eq:69}
\text{log}~Z(\alpha,\beta) = -\int_\frac{\alpha}{\beta}^\infty\!\text{d}\epsilon~g(\epsilon)\text{log}\!\left[1-\text{exp}(\alpha-\beta\epsilon)\right]
\end{equation}
due to a divergency of the logarithm at the lower bound of integration, a signature of the physical mechanism of Bose-Einstein condensation \cite{BOOK:10}. 
It is therefore not possible to proceed as in the derivation of the fermionic Bethe approximation by expanding the SP level density around this energy, which was previously associated with the Fermi energy.

This problem can be circumvented for systems with power-law SP level densities. 
As demonstrated in Ref.~\cite{Comtet2007}, for these kind of systems the bosonic MBDOS can be obtained without using the Sommerfeld expansion. 
(An analogous strategy works also for fermionic gas systems \cite{QuirinJPA}.)
Again, the key step is a saddle-point approximation of the integral Eq.~\eqref{eq:26} with the bosonic in place of the fermionic grand-canonical partition function. 
The saddle-point condition, $\Phi'\!\left(\alpha_0,\beta_0\right) = 0$, yields
\begin{equation}
N = \int_0^\infty\!\text{d}\epsilon~g(\epsilon)n^{\rm B}_{\alpha_0,\beta_0}(\epsilon) 
\quad \mbox{and} \quad 
E = \int_0^\infty\!\text{d}\epsilon~g(\epsilon)n^{\rm B}_{\alpha_0,\beta_0}(\epsilon)\epsilon
\end{equation}
where $n^{\rm B}_{\alpha_0,\beta_0}(\epsilon)$ is the Bose-Einstein distribution, namely
\begin{equation}
n^{\rm B}_{\alpha_0,\beta_0}(\epsilon) = \frac{1}{\text{exp}\!\left(-\alpha_0+\beta_0\epsilon\right)-1}.
\end{equation}

Further analytical progress is possible for power-law (smooth) SP level densities
\begin{equation}
g(\epsilon) = c\epsilon^n
\end{equation}
with $c > 0$ and $n>-1$.
Using the Bose-Einstein integral \cite{TECHREPORT:1}
\begin{equation}
\int_0^\infty\!\text{d}x~\frac{x^n}{\text{exp}(x-a)-1} = \Gamma(n+1)\text{Li}_{n+1}\!\left(e^a\right),
\end{equation}
valid for $n > -1$, one obtains 
\begin{equation}
\label{eq:70}
N = c\frac{\Gamma(n+1)}{\beta_0^{n+1}}\text{Li}_{n+1}\!\left(e^{\alpha_0}\right) \\
\quad \mbox{and} \quad 
E = c\frac{\Gamma(n+2)}{\beta_0^{n+2}}\Gamma(n+2)\text{Li}_{n+2}\!\left(e^{\alpha_0}\right).
\end{equation}
Since the physical quantities $N$ and $E$ are real-valued, Eq.~\eqref{eq:70} requires $\alpha_0 < 0$. This ensures, a posteriori, the well-behavedness of the integral Eq.~\eqref{eq:69} in a vicinity of the stationary point.

In contrast to the fermionic case \cite{QuirinJPA}, it is not possible to continue with an asymptotic expansion of the polylogarithm because now $\text{exp}\!\left(\alpha_0\right) < 1$. Instead, following Ref.~\cite{Comtet2007}, let $z = \text{exp}\!\left(\alpha_0\right)$, $0 < z < 1$, and consider two extreme limiting cases.

In the limit $z \rightarrow 1_-$, the entropy, Eq.~\eqref{eq:entropy}, at the stationary point is given by
\begin{equation}
\Phi\!\left(0,\beta_0\right) = \frac{n+2}{n+1}\beta_0E.
\end{equation}
Solving Eq.~\eqref{eq:70} for $\beta_0$ yields
\begin{equation}
\label{eq:72}
\beta_0 = \left(\frac{\theta}{E}\right)^\frac{1}{n+2},
\end{equation}
where $\theta = c\Gamma(n+2)\text{Li}_{n+2}(1)$, so that
\begin{equation}
\text{exp}\!\left(\Phi\!\left(0,\beta_0\right)\right) = \text{exp}\!\left(\frac{n+2}{n+1}\left(\theta E^{n+1}\right)^\frac{1}{n+2}\right).
\end{equation}
For example, for a gas with equally spaced SP spectra, like a collection of harmonic oscillators, one sets $c = (\hbar\omega)^{-1}$ and $n = 0$ to write
\begin{equation}
\label{eq:108}
\text{exp}\!\left(\Phi\!\left(0,\beta_0\right)\right) = \text{exp}\!\left[\pi\sqrt{\frac{2}{3}\frac{E}{\hbar\omega}}\right].
\end{equation}
In this case, since \cite{TECHREPORT:1}
\begin{equation}
\lim\limits_{z \rightarrow 1_-} \text{Li}_1(z) = \lim\limits_{z \rightarrow 1_-} -\text{log}(1-z) = \infty,
\end{equation}
the limit $z \rightarrow 1_-$ at fixed temperature, or by Eq.~\eqref{eq:72} equivalently for fixed energy $E$, corresponds to $N \to \infty$ in Eq.~\eqref{eq:70}.
Note that Eq.~\eqref{eq:108} coincides with the exponent of the fermionic Bethe approximation for the harmonic oscillator. 
This remarkable result for the MBDOS of a bosonic gas with equally spaced SP levels, valid for $1 \ll \frac{E}{\hbar\omega} \leq N$, is found in Sec.~\ref{sec:Bethe-limit} by combinatorial means.

For $|z| \ll 1$, $\text{Li}_n(z) \approx z$ \cite{TECHREPORT:1}, so that in the limit $z \rightarrow 0_+$,
\begin{equation}
\label{eq:73}
N = \frac{c}{\beta_0^{n+1}}\Gamma(n+1)z \qquad \mbox{and} \qquad 
E = \frac{c}{\beta_0^{n+2}}\Gamma(n+2)z
\end{equation}
and one obtains the (equipartition) relation
\begin{equation}
\label{eq:74}
\frac{E}{N} = \frac{n+1}{\beta_0}.
\end{equation}
Furthermore, recalling that $\alpha_0 = \text{log}(z)$, Eqs. \eqref{eq:73} and \eqref{eq:74} yield
\begin{equation}
\alpha_0 = \text{log}\!\left(\frac{(n+1)^{n+1}}{\Gamma(n+1)}\frac{N^{n+2}}{cE^{n+1}}\right)
\end{equation}
and with Eq.~\eqref{eq:entropy}, it follows that
\begin{equation}
\label{eq:75}
\text{exp}\!\left(\Phi\!\left(\alpha_0,\beta_0\right)\right) = \left(\frac{\Gamma(n+1)}{(n+1)^{n+1}}\frac{cE^{n+1}}{N^{n+2}}\right)^N\text{exp}\!\left((n+2)N\right).
\end{equation}
For fixed particle number $N$, by Eq.~\eqref{eq:73}, the limit $z \rightarrow 0_+$ corresponds to the large-temperature and hence by Eq.~\eqref{eq:74} to the large-energy limit. 

Returning to the example of the MB harmonic oscillator $c = (\hbar\omega)^{-1}$, $n = 0$, Eq.~\eqref{eq:74} is the classical equipartition of energy and Eq.~\eqref{eq:75} becomes
\begin{equation}
\text{exp}\!\left(\Phi\!\left(\alpha_0,\beta_0\right)\right) = \left(\frac{\text{exp}(N)}{N^N}\right)^2  \left(\frac{E}{\hbar\omega}\right)^N,
\end{equation}
which by the Stirling expansion of $\Gamma(N)$ \cite{BOOK:2} is for not too small $N$ approximately given by
\begin{equation}
\text{exp}\!\left(\Phi\!\left(\alpha_0,\beta_0\right)\right) \approx \frac{2\pi E}{\hbar\omega}\frac{1}{N!(N-1)!}\left(\frac{E}{\hbar\omega}\right)^{N-1}.
\end{equation}
Up to the factor $2\pi E$, probably due to the absence of the prefactor from the saddle-point approximation (which is not computed in Ref.~\cite{Comtet2007}), this reproduces the large-energy behaviour of the exact MBDOS of the bosonic harmonic oscillator determined combinatorially in Sec.~\ref{sec:combinatorial}, compare Eq.~\eqref{eq:*}.

The behaviour of the bosonic MBDOS for intermediate values of $z$ is more complex. A nice discussion can be found in Ref.\ \cite{Comtet2007}. 
It is also worth mentioning that the approach presented above can be extended beyond the saddle-point approximation \cite{Ribeiro2015} using a uniform approximation \cite{Holthaus1999}.

In summary, the key difference between the Bethe approximation and its bosonic counterpart manifests itself in the fact that the chemical potential at the stationary point of the integral Eq.~\eqref{eq:26}
equals the Fermi energy, $\epsilon_F = \alpha_0/\beta_0$, in the former case, whereas $\alpha_0/\beta_0 < 0$ in the latter case. 
While it is still possible to obtain approximations of the bosonic MBDOS by modifying the derivation of the fermionic Bethe approximation accordingly, the results are of greater analytical complexity.
Seemingly, they cover a broad parameter extent, however the range of validity of the saddle-point approximation has not been yet taken into consideration.

\section{MBDOS of systems with constant SPDOS}
\label{sec:combinatorial}

Let us reformulate the problem in terms of a combinatorial analysis.
For a MB system composed of $N$ non-interacting quantum particles, the Hamiltonian $\mathcal{H}$ is defined on the tensor product of the SP Hilbert spaces and decomposes into a sum
\begin{equation}
\mathcal{H} = \sum_{i=1}^N \mathcal{H}_i,
\end{equation}
where
\begin{equation}
\mathcal{H}_i = \mathcal{I} \otimes ... \otimes \mathcal{H}^\text{SP}_i \otimes ... \otimes \mathcal{I}
\end{equation}
operates as the SP Hamiltonian $\mathcal{H}^\text{SP}_i$ in the $i$th spot and as the identity else. We are interested in the solution of its eigenvalue problem when restricted to the subspace appropriate in the physical context. As a consequence, the MB energies are sums of SP energies and the degeneracy of a MB level can by computed by counting the number of ways in which the SP energies add up to the given MB energy and which correspond to a physically admissible state.

For the sake of concreteness, let us consider $N$ non-interacting quantum particles in one spatial dimension subject to an external harmonic potential.
No spin degrees of freedom are considered and the quantum particles are thought of as spinless, which is possible for example if they are spin-polarized, see ref.~\cite{PHDTHESIS:1} for a related discussion.
The SP energy spectra consist of equally spaced levels with spacing $\hbar\omega$, where $\hbar$ is the reduced Planck constant and $\omega$ is the harmonic potential oscillator frequency, namely we have
\begin{equation}
\left\{\epsilon^{(i)}_m = \left(m + \frac{1}{2}\right)\hbar\omega \mid m \in \mathbb{Z}_{\geq 0}\right\}
\end{equation}
for the spectrum of the $i$th oscillator, $1 \leq i \leq N$. From the general considerations above, it follows that the full MB spectrum is given by
\begin{equation}
\left\{E_{\left(m_1,...,m_N\right)} = \sum_{i=1}^N \epsilon^{(i)}_{m_i} = \left(\sum_{i=1}^N m_i\right)\hbar\omega + \frac{N}{2}\hbar\omega \mid m_1,...,m_N \in \mathbb{Z}_{\geq 0}\right\}.
\end{equation}
The ubiquitous contribution $\frac{N}{2}\hbar\omega$ is referred to as the zero-point energy of the MB harmonic oscillator. 

Now, the MBDOS reads
\begin{equation}
\label{eq:added}
\rho(N,E) = \sum_{E_{\left(m_i\right)}}g\!\left(N,E_{\left(m_i\right)}\right)\delta\!\left(E-E_{\left(m_i\right)}\right),
\end{equation}
where the sum runs over distinct MB energies and the prefactor $g\!\left(N,E_{\left(m_i\right)}\right)$ is the degeneracy of the level $E_{\left(m_i\right)}$.
Knowing the spectrum, determining the MBDOS amounts to computing the degeneracies by counting the number of independent states for which the SP energies add up to the given MB energy. This reveals the combinatorial nature of the problem. Depending on the permutation symmetry of the quantum particles, there are different answers:

Fix a MB energy $E$ and let
\begin{equation}
\label{eq:1}
{\cal E} = \frac{E}{\hbar\omega} - \frac{N}{2}.
\end{equation}

If the particles are distinguishable, one asks for the number of ordered tuples $\left(m_1,...,m_N\right)$ with $m_1,...,m_N \in \mathbb{Z}_{\geq 0}$ satisfying
\begin{equation}
\label{eq:2}
\sum_{i=1}^N m_i = {\cal E}.
\end{equation}
This number is given by
\begin{equation}
\label{eq:3}
\binom{N+{\cal E} -1}{{\cal E}}.
\end{equation}

Quantum indistinguishability adds a layer of complexity:

In the case of bosonic particles, the counting problem is solved by the number of multisets $\left[m_1,...,m_N\right]$ with entries satisfying Eq.~\eqref{eq:2}. 
Here, it is necessary to consider multisets rather than sets because multiple bosons can occupy the same SP energy level and one has to keep record of the occupation numbers, that is the multiplicities of the multiset entries. 
On the one hand, one no longer asks for ordered tuples which is an overcount due to the additional permutation symmetry. 
On the other hand, simply dividing Eq.~\eqref{eq:3} by $N!$ undercounts because it does not take into account the multiplicities of the $m_i$. Instead, the sought number is the number of partitions of ${\cal E}$ into at most $N$ positive integer parts, denoted as
\begin{equation}
\label{eq:105}
p_{\leq N}({\cal E}).
\end{equation}

\begin{figure}[t!]
\centering
\includegraphics[width=0.65\linewidth]{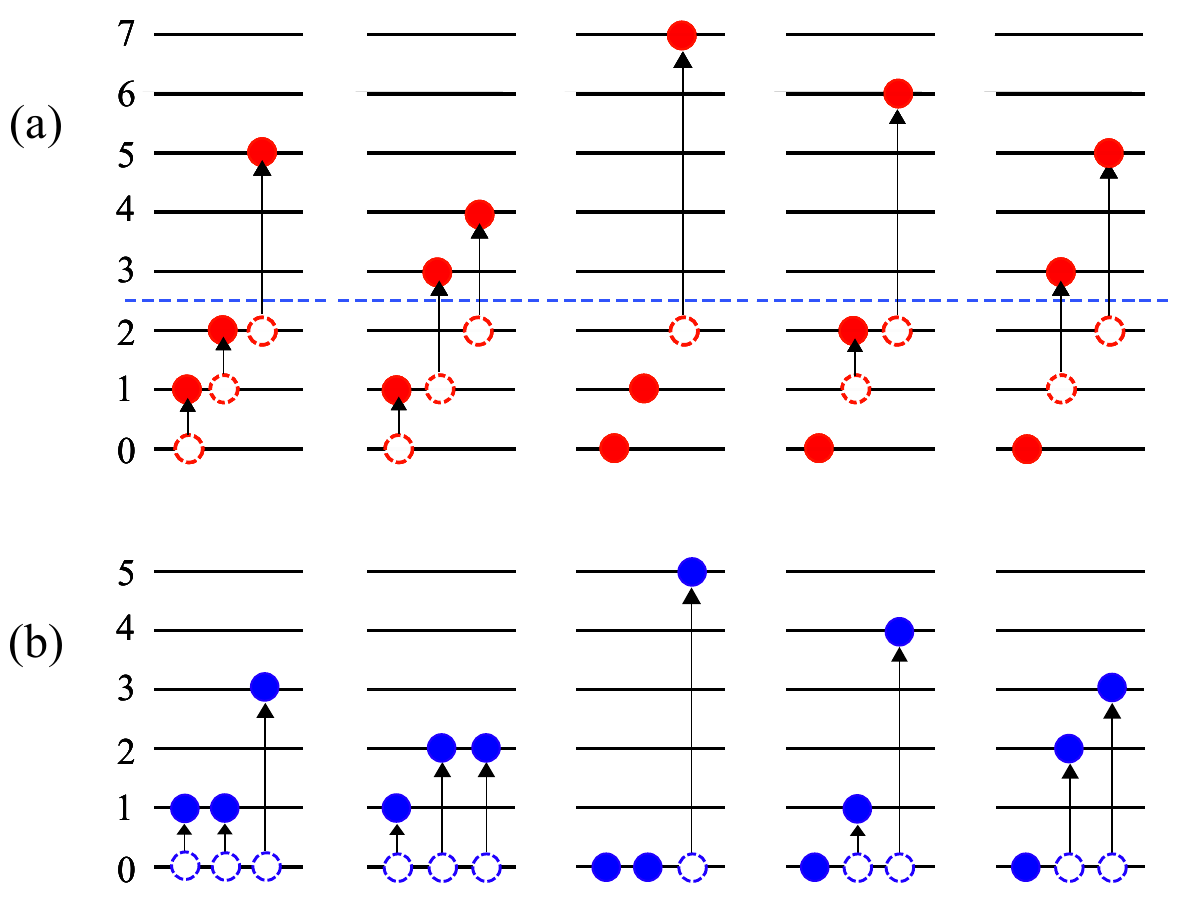}
\caption{Fermionic (a) and bosonic (b) MB configurations of $N$ harmonic oscillators at excitation energy $\hbar\omega{\cal Q}$, here $N = 3$ and ${\cal Q} = 5$. Each pannel represents, within the occupation number picture, a state of the correct symmetry with SP energies $\left(m_i + \frac{1}{2}\right)\hbar\omega$, $1 \leq i \leq 3$, indicated by filled circles, and subject to the constraint (a) $\sum_{i=1}^3 m_i = 5 + (0+1+2) = 8$, (b) $\sum_{i=1}^3 m_i = 5$. The number of independent admissible states is seen to be given by (a) $d_{=3}(8) + d_{=2}(8) = 2 + 3 = 5$, (b) $p_{\leq 3}(5) = 5$. The horizontal blue dashed line (in (a)) stands for the Fermi energy level. The vertical arrows display the SP excitations with respect to the system ground state, indicated by dashed circles. They illustrate the one-to-one mapping between fermionic and bosonic configurations, thereby demonstrating that the MB level degeneracies are the same for the two particle species.}
\label{fig:1}
\end{figure}

In the case of fermionic particles, the Pauli exclusion principle imposes the additional constraint that the $m_i$ be distinct. In this case, the combinatorial problem is solved by the number of partitions of ${\cal E}$ into exactly $N$ or $N-1$ distinct positive integer parts, denoted $d_{=N}({\cal E})$ and $d_{=N-1}({\cal E})$ respectively, hence the sum
\begin{equation}
\label{eq:4}
d_{=N}({\cal E}) + d_{=N-1}({\cal E}).
\end{equation}
Eq.~\eqref{eq:4} counts the number of multisets with distinct entries of which none or precisely one is zero, corresponding to whether or not the ground level of the physical system is occupied. 
This is illustrated in Fig.~\ref{fig:1} for the specific case of $N=3$ and an excitation energy $Q=\hbar \omega {\cal Q}$ with ${\cal Q} = 5$.

These results for the degeneracies are found to be stated in Ref.\ \cite{Auluck1946} among others. 
The discrepancy by a factor of $N!$ between the results for distinguishable particles given in the reference and those of Eq.~\eqref{eq:3} will be addressed in Sec.~\ref{sec:distinguishability}.

It is illustrative to recast Eq.~\eqref{eq:4} as follows. Write
\begin{equation}
{\cal E} = {\cal Q}^{(f)} + {\cal E}_\text{GS}^{(f)}(N),
\end{equation}
where ${\cal E}_\text{GS}^{(f)}(N)$ denotes the fermionic MB ground-state energy in units of $\hbar\omega$, not including zero-point energy, namely
\begin{equation}
{\cal E}_\text{GS}^{(f)}(N) = \sum_{m=0}^{N-1} m = \frac{N(N-1)}{2}.
\end{equation}
After substituting the above relation into Eq.~\eqref{eq:4}, a sequence of combinatorial manipulations involving the identities (obtained similarly as in the proof of \cite{BOOK:2}, 2.5, Theorem C)
\footnote{See also entries A008289 and A026820 in \it{The On-Line Encyclopedia of Integer Sequences}, \url{https://oeis.org}, accessed December 2022.}
\begin{equation}
\label{eq:5}
d_{=N}(x) = p_{=N}\!\left(x-\binom{N}{2}\right)
\end{equation}
and
\begin{equation}
\label{eq:6}
p_{=N}(x) = p_{\leq N}(x-N),
\end{equation}
with $p_{=N}(x)$ the number of partitions of $x$ into exactly $N$, but not necessarily distinct positive integer parts, yields
\begin{equation}
\label{eq:106}
\begin{split}
&d_{=N}\!\left({\cal Q}^{(f)}+\frac{N(N-1)}{2}\right) + d_{=N-1}\!\left({\cal Q}^{(f)}+\frac{N(N-1)}{2}\right) \\
&\stackrel{\eqref{eq:5}}{=} p_{=N}\!\left({\cal Q}^{(f)}+\frac{N(N-1)}{2}-\binom{N}{2}\right) + p_{=N-1}\!\left({\cal Q}^{(f)}+\frac{N(N-1)}{2}-\binom{N-1}{2}\right) \\
&= p_{=N}\!\left({\cal Q}^{(f)}\right) + p_{=N-1}\!\left({\cal Q}^{(f)}+(N-1)\right) \\
&\stackrel{\eqref{eq:6}}{=} p_{=N}\!\left({\cal Q}^{(f)}\right) + p_{\leq N-1}\!\left({\cal Q}^{(f)}\right) \\
&= p_{\leq N}\!\left({\cal Q}^{(f)}\right).
\end{split}
\end{equation}
So
\begin{equation}
d_{=N}({\cal E}) + d_{=N-1}({\cal E}) = p_{\leq N}\!\left({\cal Q}^{(f)}\right).
\end{equation}
Observing that for bosons ${\cal E}_\text{GS}^{(b)}(N) = 0$, one can summarize the results for the degeneracies as
\begin{equation}
\label{eq:7}
\begin{split}
&g^{(d)}(N,E) = \binom{N+{\cal E}-1}{\cal E} \\
&g^{(b)}(N,E) = p_{\leq N}\!\left({\cal E} - {\cal E}_\text{GS}^{(b)}(N)\right) = p_{\leq N}\!\left({\cal Q}^{(b)}\right) \\
&g^{(f)}(N,E) = p_{\leq N}\!\left({\cal E} - {\cal E}_\text{GS}^{(f)}(N)\right) = p_{\leq N}\!\left({\cal Q}^{(f)}\right)
\end{split}
\end{equation}
where $d$, $b$ and $f$ denote distinguishable particles, bosons and fermions respectively, and we recall
\begin{equation}
\label{eq:8}
E = \hbar\omega\left({\cal E} + \frac{N}{2}\right) = \hbar\omega {\cal Q}^{(f,b)} + \hbar\omega\left({\cal E}_\text{GS}^{(f,b)}(N) + \frac{N}{2}\right) = Q^{(f,b)} + E_\text{GS}^{(f,b)}(N).
\end{equation}
In particular, the degeneracies are the same for the two species of indistinguishable particles up to a shift in the energy argument which corresponds to the difference between the fermionic and bosonic ground-state energies
\begin{equation}
E_\text{GS}^{(f)}(N) - E_\text{GS}^{(b)}(N) = \hbar\omega\frac{N(N-1)}{2}.
\end{equation}

Figure~\ref{fig:1} explains why this observation is consistent.
Formally, the figure gives a particular bijection between the set of partitions of an integer ${\cal Q}$ into at most $N$ parts and the set of partitions of the integer ${\cal Q} + \frac{N(N-1)}{2}$ into $N$ or $N-1$ distinct parts. 
Namely, given a partition of ${\cal Q}$ into at most $N$ parts, say ${\cal Q} = x_1 + ... + x_N$ with $0 \leq x_1 \leq ... \leq x_N$, obtain a partition ${\cal Q} + \frac{N(N-1)}{2} = \left(x_1 + 0\right) + \left(x_2 + 1\right) + ... + \left(x_N + (N-1)\right)$ into exactly $N$ or $N-1$ distinct parts, which can be thought of as distributing $x_1, ..., x_N$ quanta of excitation energy to $N$ fermions residing in the ground state. 
In short, both the bosonic and fermionic MB states are attained by dividing the number of excitation quanta among the quantum particles in the respective ground state. 

While related considerations can be found in Refs.~\cite{Leboeuf2005b, Leboeuf2005, MISC:2}, the conclusions presented here are derived independently, emphasizing the combinatorial perspective. In particular, the results in Eq.~\eqref{eq:7} are established by rigorously bridging the gap between the earlier literature \cite{Auluck1946} and the later references through Eq.~\eqref{eq:106}.

Note that the zero-point energy of the oscillators enters in the degeneracies Eq.~\eqref{eq:7} only as a shift in the argument of the combinatorial function and, hence, does not affect the combinatorics. 
Therefore, the results generalize to SP spectra of the form 
$\left\{\epsilon_m = (m + c)\Delta \mid m \in \mathbb{Z}_{\geq 0}\right\}$
for any constant $c$ and unit of energy $\Delta$, with the only modification that Eq.~\eqref{eq:1} is replaced by ${\cal E} = \frac{E}{\Delta} - cN$.

\subsection{Bounds on $p_{\leq N}({\cal Q})$}
\label{sec:bounds}

For the computation of the number of partitions $p_{\leq N}({\cal Q})$, recursive formulas as well as series expansions are available, some of which are reviewed in ref.\ \cite{BOOK:2}. 
However, explicit bounds on $p_{\leq N}({\cal Q})$ can be derived. By eq. \eqref{eq:106}, one has
\begin{equation}
p_{\leq N}({\cal Q}) = d_{=N}\!\left({\cal Q}+\frac{N(N-1)}{2}\right) + d_{=N-1}\!\left({\cal Q}+\frac{N(N-1)}{2}\right)
\end{equation}
which equals the cardinality of the subset of
\begin{equation}
X = \left\{\left(x_i\right)_{1 \leq i \leq N} \mid x_i \in \mathbb{Z}_{\geq 0}, \sum_i x_i = {\cal Q} + \frac{N(N-1)}{2}\right\}
\end{equation}
that consists of the tuples with distinct entries (of which one may be zero), divided by $N!$ since the order of the parts is irrelevant. Therefore,
\begin{equation}
p_{\leq N}({\cal Q}) \leq \frac{|X|}{N!} \leq \frac{1}{N!}\binom{N+{\cal Q}+\frac{N(N-1)}{2}-1}{{\cal Q}+\frac{N(N-1)}{2}} = \frac{1}{N!}\binom{{\cal Q}+\frac{N(N+1)}{2}-1}{N-1}.
\end{equation}
On the other hand, it has already been argued that (see text above Eq.~\eqref{eq:105})
\begin{equation}
p_{\leq N}({\cal Q}) \geq \frac{1}{N!}\binom{N+{\cal Q}-1}{{\cal Q}} = \frac{1}{N!}\binom{{\cal Q}+N-1}{N-1},
\end{equation}
so that in summary
\begin{equation}
B_1 = \frac{1}{N!}\binom{{\cal Q}+N-1}{N-1} \leq p_{\leq N}({\cal Q}) \leq \frac{1}{N!}\binom{{\cal Q}+\frac{N(N+1)}{2}-1}{N-1} = B_2.
\end{equation}
These bounds are also derived in ref.\ \cite{Agnaesson2002} in a more general framework. Though they miscount considerably as $N$ increases, they are asymptotically tight in ${\cal Q}$ in the sense that
\begin{equation}
\lim\limits_{Q \to \infty} \frac{B_1}{B_2} = 1.
\end{equation}

\subsection{Bethe approximation and other limits}
\label{sec:Bethe-limit}

Here, we relate the exact MB energy level degeneracies summarized in Eq.~\eqref{eq:7} to the Bethe approximation discussed in Sec.~\ref{sec:background}.

As long as ${\cal Q} \leq N$, bounding the number of parts of a partition of ${\cal Q}$ by $N$ does not impose any restrictions, so $p_{\leq N}({\cal Q}) = p({\cal Q})$, where $p({\cal Q})$ denotes the number of integer partitions of ${\cal Q}$ with an arbitrary number of parts.   
The asymptotic limit, ${\cal Q} \gg 1$, of $p({\cal Q})$ is well-known in the mathematical literature \cite{Hardy1918}, namely \footnote{A series expansion of $p({\cal Q})$ was later obtained by Rademacher \cite{Rademacher1937, Rademacher1938}.}
\begin{equation}
\label{eq:10}
p({\cal Q}) \sim \frac{1}{4\sqrt{3}{\cal Q}}\text{exp}\!\left(\pi\sqrt{\frac{2{\cal Q}}{3}}\right) \,.
\end{equation}
Using Eq.~\eqref{eq:8} and taking the mean SP level density at the Fermi energy to be $g\!\left(\epsilon_F\right) = (\hbar\omega)^{-1}$, Eq.~\eqref{eq:10} is precisely the Bethe approximation for the fermionic MB harmonic oscillator given by Eq.~\eqref{eq:38}.
The relation between $p({\cal Q})$ and the Bethe approximation is also remarked in Refs.~\cite{Leboeuf2005, Leboeuf2005b}. 

As a consequence, in the case of the MB harmonic oscillator, the Bethe approximation holds for bosons as well since the degeneracies are likewise given by $p_{\leq N}({\cal Q})$. This is noteworthy because in general, as was discussed in Sec.~\ref{sec:Bethe-bosonic}, the derivation of the Bethe approximation does not immediately translate to the bosonic case. 

The Bethe formula, Eq.~\eqref{eq:38}, depends on the number of particles only through $g\!\left(\epsilon_F\right)$ and this dependence vanishes for the harmonic oscillator. 
This can be understood from the combinatorial viewpoint because even at an exact level, $p_{\leq N}({\cal Q}) = p({\cal Q})$ as long as ${\cal Q} \leq N$, so in particular within the range of validity of the Bethe approximation. 
This means that when fixing ${\cal Q} \leq N$ and formally adding a particle to the system, the degeneracies of both fermionic and bosonic MB harmonic oscillators do not change. 
The combinatorial reason for this is that already before adding a particle, all possible ways of distributing the available ${\cal Q}$ excitation quanta to any number of indistinguishable entities $\geq {\cal Q}$ had been exhausted. 
Note that in the fermionic case, fixing the excitation energy and adding a particle entails increasing the ground-state energy and thereby the total energy. 
However, the reasoning remains the same.
Effectively, the Fermi sea is increased by one particle while all individual excitation quanta remain the same.
In contrast, adding a particle in the distinguishable case gives rise to new MB states due to the lack of permutation symmetry.

\subsection{Distinguishable particles and the correct Boltzmann counting}
\label{sec:distinguishability}

Using that in the large-energy limit \cite{Erdös1941},
\begin{equation}
\label{eq:*}
\lim\limits_{{\cal Q} \to \infty} \frac{\binom{{\cal Q}-1}{N-1}}{N! \, p_{\leq N}({\cal Q})} = 1,
\end{equation}
one has for the degeneracies
\begin{equation}
\lim\limits_{{\cal Q} \to \infty} \frac{\binom{N+{\cal Q}-1}{{\cal Q}}}{N! \cdot p_{\leq N}({\cal Q})} = \lim\limits_{{\cal Q} \to \infty} \frac{\binom{N+{\cal Q}-1}{N-1}}{\binom{{\cal Q}-1}{N-1}} = 1,
\end{equation}
that is
\begin{equation}
g^{(d)}(N,Q) \sim N! \, g^{(f,b)}\!\left(N,Q + E_\text{GS}^{(f,b)}(N)\right)
\end{equation}
asymptotically in the energy. In this limit, indistinguishable and distinguishable particles hence compare as in the classical case, where in place of the degeneracies, it is the classical phase-space volume which is larger by a factor $N!$ for distinguishable than for indistinguishable particles,
in analogy to the so-called ``correct Boltzmann counting" discussed in the statistical mechanics \cite{Jaynes1992, INCOLLECTION:1}.
In Ref.~\cite{Auluck1946}, this factor is included in the degeneracies in order that they agree asymptotically.

This limiting behaviour, too, can be understood combinatorially by considering
\begin{equation}
\label{eq:85}
\lim\limits_{{\cal Q} \to \infty} \frac{d_{=N}({\cal Q})}{p_{\leq N}({\cal Q})} = \lim\limits_{{\cal Q} \to \infty} \frac{p_{\leq N}\!\left({\cal Q}-\frac{N(N+1)}{2}\right)}{p_{\leq N}({\cal Q})} 
= \lim\limits_{{\cal Q} \to \infty} \frac{\binom{{\cal Q}-\frac{N(N+1)}{2}-1}{N-1}}{\binom{{\cal Q}-1}{N-1}} = 1.
\end{equation}
For large ${\cal Q}$, the number of partitions of ${\cal Q}$ into at most $N$ parts is thus dominated by the number of distinct partitions of ${\cal Q}$ into exactly $N$ parts and each of the corresponding MB configurations of indistinguishable particles gives, upon permuting the particles, $N!$ many MB configurations of distinguishable particles.

\subsection{Prospects of generalization}
\label{sec:generalization}

While the discussion at the beginning of the present section, leading to Eq.~\eqref{eq:added}, applies to an arbitrary non-interacting MB system, determining the MB level degeneracies is, in general, a difficult combinatorial problem. The case of constant SPDOS considered in this work has the virtue that the counting problem becomes tractable, and a natural question is to what extent the combinatorial approach and its results can be extended to other physical systems.

For $N$ particles under harmonic confinement in more than one, say $D$, spatial dimensions, computing the degeneracies remains a linear integer partition problem, admitting a similar analysis as conducted here. For distinguishable particles, the system is equivalently described as an assembly of $ND$ one-dimensional harmonic oscillators and therefore covered by our results Eq.~\eqref{eq:7}, whereas for indistinguishable particles, modified constraints on the parts of the integer partition have to be taken into account.

An energy-dependent mean SP level spacing, in contrast, adds considerably to the complexity of the counting problem. Consequences of fluctuations about a rigid spectrum are investigated in Refs.~\cite{Leboeuf2005b, MISC:5}. For SP energies obeying a power law in the quantum number, the problem of computing the MB level degeneracies translates into a non-linear partition problem, that is the counting of partitions into powers of integers, for which asymptotic results are obtained in Ref.~\cite{Agarwala1951}. We remark that the finding to which the results of Ref.~\cite{Agarwala1951} specialize in the case of the harmonic oscillator and which in our notation reads $d_{=N}\!\left({\cal Q}+\frac{N^2}{2}\right) \sim p_{\leq N}({\cal Q})$ for large $N$ and ${\cal Q}$, can also be deduced from Eq.~\eqref{eq:85}.

\section{Summary and conclusions}
\label{sec:conclusion}

The mean level density is a fundamental quantity in the analysis of the properties of quantum systems since it determines their natural energy scale.
Its precise knowledge has been a longstanding challenge in quantum physics, in particular for many-body systems.
The study of the MBDOS has a particularly rich history in nuclear physics, modeled as fermionic gas systems \cite{Bethe1936, BohrMottelson1998} and, more recently, in bosonic gases \cite{Comtet2007}.
These studies have employed ingenious approximation strategies to expand both the fermionic and bosonic partition function, leading to closed-form expressions for the MDDOS.

There is an alternative line of thinking where one investigates the MBDOS of systems of non-interacting quantum particles by means of combinatorial approaches, which make it possible to derive exact solutions in the simplified setting of equally-spaced single-particle spectra \cite{Auluck1946, Leboeuf2005b, Leboeuf2005}. In the present work, we advance this combinatorial analysis, extending it beyond asymptotic regimes and promoting the comparison of the different ensembles. Thereby, we rigorously bridge the gap between the results in the existing literature. The expressions for the MBDOS that we establish in this framework are exact and have no limitation to their range of validity.

A remarkable finding is the emergence of a mapping between the excitation spectra of fermion and boson systems with constant SP level spacing. This mapping reveals that the MBDOS for both particle types coincide, up to a shift due to ground state energy.

Furthermore, our combinatorial analysis yields another nice result: The MBDOS is independent of the number of particles $N$ within a specific range. This interval of $N$-independence comprises the validity range of the Bethe approximation, giving further support for its accuracy in this context.

\section*{Acknowledgments}
We thank Quirin Hummel for useful conversations.
CL thanks the Brazilian funding agencies CNPq and FAPERJ. GM is funded through a fellowship by the Studienstiftung des Deutschen Volkes. We acknowledge financial support from the Deutsche Forschungsgemeinschaft (German Research Foundation) through Ri681/15-1 (project number 456449460) within the Reinhart-Koselleck Programme.
Finally, we would like to thank the anonymous referees for their comments, which led to improvements of the manuscript and prompted us to include subsection \ref{sec:generalization}.

\section*{References}
\bibliography{mbdos_of_bosonic_and_fermionic_gases}

\end{document}